\documentclass[iop,apj]{emulateapj}

\newcommand{\myemail}{imai@kugi.kyoto-u.ac.jp}

\slugcomment{Submitted to ApJ on 2016 February 21}

\shorttitle{Beaming structures of Jovian S-bursts}
\shortauthors{Imai et al.}

\begin{document}


\title{Beaming structures of Jupiter's decametric common S-bursts observed from LWA1, NDA, and URAN2 radio telescopes}


\author{Masafumi Imai\altaffilmark{1}, Alain Lecacheux\altaffilmark{2}, Tracy E. Clarke\altaffilmark{3}, Charles A. Higgins\altaffilmark{4}, \\ Mykhaylo Panchenko\altaffilmark{5}, Jayce Dowell\altaffilmark{6}, Kazumasa Imai\altaffilmark{7}, Anatolii I. Brazhenko\altaffilmark{8}, \\ Anatolii V. Frantsuzenko\altaffilmark{8}}
\author{Alexandr A. Konovalenko\altaffilmark{9}}

\altaffiltext{1}{Department of Geophysics, Kyoto University, Kyoto 606-8502, Japan}

\altaffiltext{2}{Laboratoire d'Etudes Spatiales et d'Instrumentation en Astrophysique, CNRS/Observatoire de Paris, Meudon F-92195, France}

\altaffiltext{3}{Naval Research Laboratory, Washington, DC 20375, USA}

\altaffiltext{4}{Department of Physics and Astronomy, Middle Tennessee State University, Murfreesboro, TN 37132, USA}

\altaffiltext{5}{Space Research Institute, Austrian Academy of Sciences, Graz A-8042, Austria}

\altaffiltext{6}{Department of Physics and Astronomy, University of New Mexico, Albuquerque, NM 87131, USA}

\altaffiltext{7}{Department of Electrical Engineering and Information Science, Kochi National College of Technology, Kochi 783-8508, Japan}

\altaffiltext{8}{Poltava Gravimetrical Observatory, S.~Subotin Institute of Geophysics, National Academy of Sciences of Ukraine, Poltava 36029, Ukraine}

\altaffiltext{9}{Institute of Radio Astronomy, National Academy of Sciences of Ukraine, Kharkiv 61002, Ukraine}


\email{Correspondence to: M. Imai, \myemail}


\begin{abstract}
On 2015 February 21, simultaneous observations of Jupiter's decametric radio emission between 10 and 33 MHz were carried out using three powerful low-frequency radio telescopes: Long Wavelength Array Station One (LWA1) in USA; Nan\c{c}ay Decameter Array (NDA) in France; and URAN2 telescope in Ukraine. We measure lag times of short-bursts (S-bursts) for 105-minutes of data over effective baselines up to 8460 km by using cross-correlation analysis of the spectrograms from each instrument. Of particular interest is the measurement of the beaming thickness of S-bursts, testing if either flashlight- or beacon-like beaming is emanating from Jupiter. We find that the lag times for all pairs drift slightly as time elapses, in agreement with expectations from the flashlight-like beaming model. This leads to a new constraint of the minimum beaming thickness of 2.66". Also, we find that most of the analyzed data abound with S-bursts, whose occurrence probability peaks at 17-18 MHz.
\end{abstract}





\section{Introduction}

Among the planetary auroral radio components in our solar system, Jupiter's decametric (DAM) radiation dominates, having the strongest flux density up to $10^{-19}$ $\mathrm{W}/(\mathrm{m}^2\cdot \mathrm{Hz})$  in a frequency range from a few to 40 MHz \citep[][and references therein]{Zarka:1998jgr,Clarke:2004ba}. Ground-based radio telescopes are limited to observations above 10 MHz due to the effects of the Earth's ionosphere. Jovian nonthermal DAM radio emission abounds with various kinds of complex frequency structures on the order of milliseconds to tens of seconds \citep{Gallet:1961ba}. These may be classified as short-bursts (abbreviated as S-bursts), which change on a time scale of milliseconds, and long-bursts (abbreviated as L-bursts), which vary with a temporal change of seconds. The latter can be widely observed in all of Jovian DAM spectrum, while the former is seen in a part of the DAM spectrum classified as Io-related DAM (Io-DAM) because of a strong influence of the Galilean moon Io \citep{Bigg:1964nat}.

Earth-based radio observatories detect the Io-DAM cone emissions \citep{Dulk:1970apj} when the cones are centered on the Jovian dawn and dusk sides; the former is further classified as Io-B and -D, and the latter as Io-A and -C. These cone emissions result from the electron cyclotron maser instability \citep{Treumann:2006aar}, where the wave-particle interaction in the strong magnetized plasma of Jupiter's poles produces the right-hand extraordinary (R-X) mode radiation near the electron gyrofrequency. Therefore, the sources of the Io-A and -B and of the Io-C and -D are believed to be located in Jovian northern and southern polar regions, respectively.

Long baseline analyses of common Jovian S-bursts have been carried out, using various baselines from 1200 to 6980 km and frequency coverages of 22 to 34 MHz \citep{Dulk:1970apj, Lynch:1976aj, Rucker:2001pre, Lecacheux:2004pss}. \citet{Dulk:1970apj} first attempted to measure the arrival time of S-bursts at 34 MHz using the two 0.31-second S-burst events from radio telescopes in Boulder, Colorado, and Clark Lake, California (i.e. 1200 km apart). It was, however, inconclusive due to a lack of recorded time accuracy. Later \citet{Lynch:1976aj} analyzed the 6-second period of a train of 68 S-bursts over a 6980 km baseline at 18 MHz, insisting that all of them were simultaneously emanated from Jupiter within a north-south view of about 1.8". In contrast to the single spectral channel monitoring, the three events of synchronized Jovian S-bursts in a few second period show good correlation in a frequency range of 22 to 26 MHz between radio telescopes in Nan\c{c}ay, France, and Kharkiv, Ukraine \citep{Rucker:2001pre, Lecacheux:2004pss}, but it was not well-discriminated between two different wavefront arrival times because of both the accuracy of temporal sampling and relatively short baseline of about 3000 km.

\begin{figure*}
 \noindent\includegraphics[width=43pc]{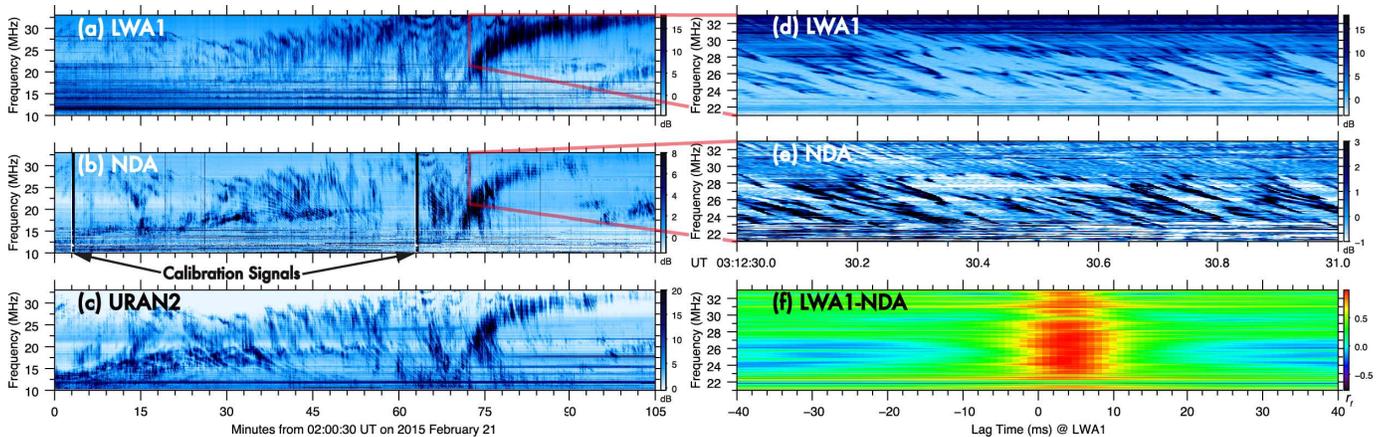}
  \caption{Coordinated LWA1-NDA-URAN2 observations of Jupiter's Io-D/B event on 21 February 2015. The frequency-time plots used in this study are depicted for (a) LWA1, (b) NDA, and (c) URAN2 radio telescopes. Also, the one-second dynamic spectra containing a train of S-bursts are displayed for (d) LWA1 and (e) NDA, and the cross-correlated spectrum is plotted for (f) LWA1-NDA.}
 \label{figs:plot_lwa-nda-uran2}
\end{figure*}

The purpose of this paper is to present the results of a detailed cross-correlation analysis of Jovian S-bursts in a 105-minute duration observation obtained using three interferometers spanning a maximum baseline of 8460 km. We measure an angular size of the beam of 2.66", thereby suggesting this cone thickness along the east-west direction. Also, new statistical S-bursts properties regarding the observed frequency of 10 to 33 MHz are investigated.

\begin{deluxetable*}{c c c c c}
\tabletypesize{\small}
\tablecolumns{5}
\tablewidth{0pt}
\tablecaption{East-West Angular Size and Effective Baseline During the Analyzed Time \label{table:obs_para}}
\tablehead{
  \colhead{Telescope Pair} & 
  \multicolumn{2}{c}{East-West Angular Size (")} & 
  \multicolumn{2}{c}{Effective Baseline (km)\,\tablenotemark{a}} \\
  \colhead{} &
  \colhead{Start (02:00:30 UT)} & 
  \colhead{Stop (03:45:30 UT)} & 
  \colhead{Start (02:00:30 UT)} & 
  \colhead{Stop (03:45:30 UT)} }
\startdata
NDA-URAN2\,\tablenotemark{b} & 0.30 & $-0.02$ & 960 & 56.9 \\
LWA1-NDA\,\tablenotemark{c} & 2.36 & 2.32 & 7500 & 7090 \\
LWA1-URAN2 & 2.66 & 2.21 & 8460 & 7040 
\enddata
\tablenotetext{a}{The effective baseline is defined as a distance projected on Jovian CML of System III.}
\tablenotetext{b}{The effective baseline and angular size are zero at 03:39:47 UT.}
\tablenotetext{c}{The maximum angular size appears as 2.37" at 02:26:04 UT, which corresponds to 7540 km baseline.}
\end{deluxetable*}

\section{Data and Observations}
The simultaneous observations of Jovian radio bursts were made at the Long Wavelength Array Station One (LWA1) on the Plains of San Agustin, New Mexico, USA, at the Nan\c{c}ay Decameter Array (NDA) in Nan\c{c}ay, France, and at the URAN2 radio telescope in Poltava, Ukraine. All of the three radio telescopes are capable of measuring the full Stokes parameters, but only the Stokes I is used in this study. A number of antennas, respectively, comprise 256 bow-tie dipoles for LWA1 \citep{Ellingson:2013apieee}, 144 conical helices for NDA \citep{Boischot:1980ica, Lecacheux:2000ba}, and 512 cross dipoles for URAN2  \citep{Brazhenko:2005ki}. The DRX mode from LWA1 is capable of recording the time-waveform at two tunings centered at 20 and 28 MHz with a total bandwidth of 19.6 MHz each \citep{Clarke:2014jgr}. Per the Fourier transform post-processing \citep{Dowell:2012jai}, the spectral data used in this study having a frequency coverage of 10.2 to 37.8 MHz with temporal resolution 4.8 ms and spectral resolution 4.79 kHz. The ``Routine" spectrometer installed at NDA provided spectrograms of the first 100 MHz frequency band, sampled every 48.83 kHz and 5 ms steps in frequency and time, respectively. The digital spectrometer DSP-Z \citep{Ryabov:2010aa} employed in URAN2 monitored the frequency coverage of 8.25 to 33 MHz with a 4.03 kHz step, measuring the sample interval of 20.1 ms. Overall, the overlapped frequency of LWA1/DRX, NDA/Routine, and URAN2/DSP-Z ranges from 10.2 to 33 MHz, which covers most of the full spectra of Jupiter's radio bursts.

Figures \ref{figs:plot_lwa-nda-uran2}a--\ref{figs:plot_lwa-nda-uran2}c show the overview of the coordinated LWA1-NDA-URAN2 observations, which consist of both L- and S-bursts with a duration of 1 hour and 45 minutes starting at 02:00:30 UT on 2015 February 21. The Jovian central meridian longitude (CML) ranged from $91.5^\circ$ to $155.0^\circ$ and Io was at $86.5^\circ$--$101.4^\circ$ from superior geocentric conjunction. Concerning the three possible telescope pairs, the NDA-URAN2, LWA1-NDA, and LWA1-URAN2 pairs, the angular size in the east-west direction and effective baseline are listed in Table \ref{table:obs_para}. In short, the effective baseline covers from 0 to 8460 km along which the east-west angular size varies as a view from Jupiter, varies from $-0.02$" to 2.66". It is obvious that the similar Jovian radio Io-D/B structures are seen for all of the radio telescopes. Note that two vertical lines in Figure \ref{figs:plot_lwa-nda-uran2}b correspond to the hourly-routine calibration signals.

\begin{figure}
 \centering
 \noindent\includegraphics[width=20pc]{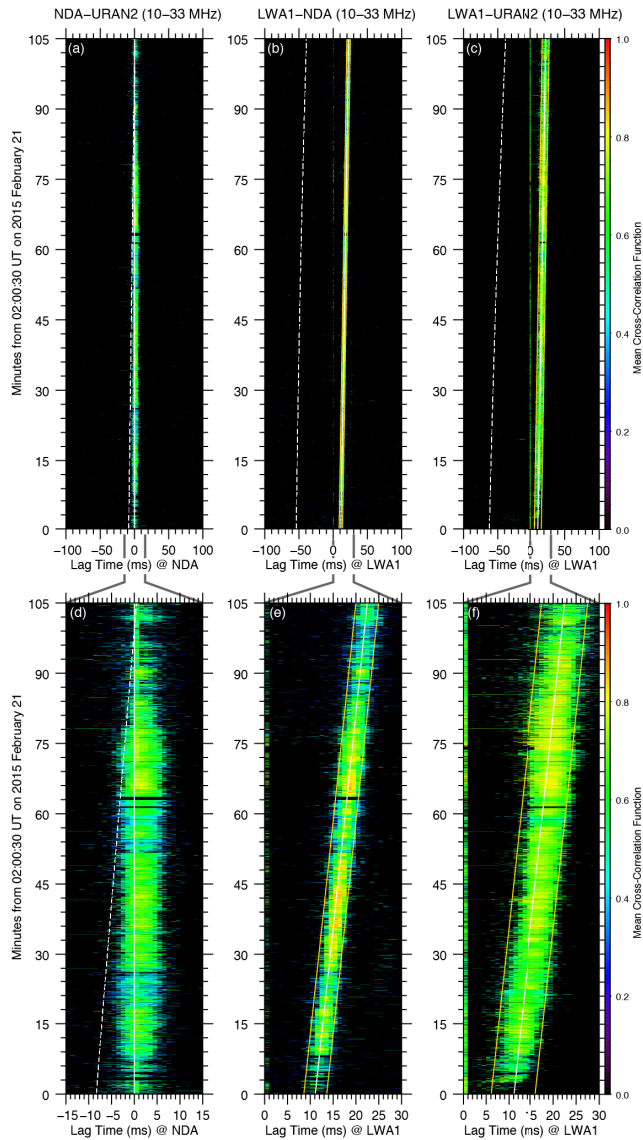}
  \caption{Mean cross-correlation functions from 10.2 to 33 MHz in a pair of (a, d) NDA-URAN2, (b, e) LWA1-NDA, and (c, f) LWA1-URAN2 plotted as a function of lag time and elapsed time from 02:00:30 UT. The white solid and dotted lines indicate the expected delay time deduced from the flashlight-like beaming and beacon-like beaming, respectively. The lag times when the high mean cross-correlation functions are exhibited are clearly fitted with the expected time for the flashlight-like beaming model. The two yellow lines are the error range of $\pm 2.5$ ms for (b, e) and of $\pm 5$ ms for (c, f).}
 \label{fig:x-lwa1-nda-uran2}
\end{figure}

\section{Analysis and Beaming}

Although the Very Long Baseline Interferometry (VLBI) method was applied to Jupiter's S-bursts \citep{Dulk:1970apj, Lynch:1976aj}, the cross-correlation analysis of the S-burst dynamic spectrum in a few second period was performed by \citet{Rucker:2001pre} and \citet{Lecacheux:2004pss}. Briefly, one of the differences is that the data format for the VLBI method are stored with waveform voltage induced from the antennas and the dynamic spectrum data are Fourier-transformed spectral data (i.e. Stokes I) derived from the VLBI data. In this study, we have extended the cross-correlation analysis to the LWA1-NDA-URAN2 spectrograms with which the sampling step varies with both temporal and spectral resolution.

In the resampling process, we average the observed values related to the Stokes I at each station over a spectral resolution of 50 kHz, and then resample the obtained values with a common fixed temporal resolution of 1 ms by means of a cubic interpolation called the Catmull-Rom spline. We compute the frequency-dependent cross-correlation function $r_f$, which can be described as
\begin{eqnarray}
r_f \left(\tau\right) &=& \frac{P_{xy}}{\displaystyle \sqrt{\sum_t \left( x_f \left(t\right)-\bar{x}_f\right)^2}\cdot \sqrt{\sum_t \left(y_f \left(t\right)-\bar{y}_f \right)^2}} \\
\mathrm{with}&& \nonumber \\
P_{xy} &=& \displaystyle \sum_t \left[ \left(x_f \left(t\right)-\bar{x}_f \right)\cdot \left( y_f \left(t+\tau\right)-\bar{y}_f\right)\right] \left(\tau \ge 0\right) \\
&=& \displaystyle \sum_t \left[ \left(x_f \left(t+|\tau|\right)-\bar{x}_f \right)\cdot \left( y_f \left(t\right)-\bar{y}_f\right)\right] \left(\tau < 0\right)
\end{eqnarray}
where the variables are defined as follows: $\tau$, shifted temporal variable (lag time); t, temporal variable in one-second period; $x_f$ and $y_f$, observed Stokes I for each frequency in dB in a pair of either NDA-URAN2, LWA1-NDA, or LWA1-URAN2; and $\bar{x}_f$ and $\bar{y}_f$, mean values of $x_f$ and $y_f$. Note that a positive $\tau$ is when the time of $x_f$ is more delayed than the time of $y_f$ and vice versa.

Measuring accurate arriving times at each station is a key to extract a nature of the S-bursts. The LWA1 and NDA recorded the timing (inserted from GPS synchronized clock time) and sampled a constant temporal cycle with an accuracy of better than 1 ms. However the URAN2's initial timing was compromised due to set-up not being on GPS but DPS-Z internal time. In order to approximately correct the URAN2's timing, we have cross-correlated the NDA and URAN2 artificial-noise-free spectrograms from 10.20 to 10.45 MHz in a 50 kHz interval. This contains approximately the same amount of propagation lag time, about a few milliseconds from both Jovian S-bursts and the sweeping signals from the European Ionosonde stations \citep{Belehaki:2006sw}. After adjusting the corrected lag time for the URAN2 data, we investigate the trend of lag times in a pair of the NDA-URAN2, LWA1-NDA, and LWA1-URAN2 in a wide frequency range of 10--33 MHz.

An example of the cross-correlation spectrum is shown in Figure \ref{figs:plot_lwa-nda-uran2}f by cross-correlating the dominant intensity of S-bursts from LWA1 in Figure \ref{figs:plot_lwa-nda-uran2}d and NDA in Figure \ref{figs:plot_lwa-nda-uran2}e. In this case, the lag time ($\tau_\mathrm{max}$) is 4 ms when the cross-correlation functions are highest. We have applied this technique to find $\tau_\mathrm{max}$, and the mean ($\mu$) and standard deviation ($\sigma$) of $r_f$ in one-second periods for each frequency; the total number of computations for $\tau_\mathrm{max}$ amount to about seventeen-billion points (more precisely, 457 spectral channels $\times$ 6300 one-second periods $\times$ 1999 one-millisecond lag time steps $\times$ 3 telescope pairs). Note that, in order to discard quasi-random background fluctuations, we keep datum points only if $r_f$ of the lag time is above the threshold of $\mu$ + $z_{99.5\%}\sigma$, where $z_{99.5\%}$ is a 99.5\% confidence interval of a Gaussian distribution.

After collecting representative lag times and corresponding cross-correlation functions for each frequency in a bin of 1 ms, we compute the mean cross-correlation functions in each bin that are defined by a sum of cross-correlation functions over a total number of observed counts. In order to remove temporal changes in data reception, we impose a condition where the observed counts in a bin are above 4 points (1\% of total spectral channels). Figure \ref{fig:x-lwa1-nda-uran2} shows a final form of the mean cross-correlation functions depicted as a function of elapsed time from 02:00:30 UT and lag time. 

\begin{figure}
 \centering
 \noindent\includegraphics[width=19pc]{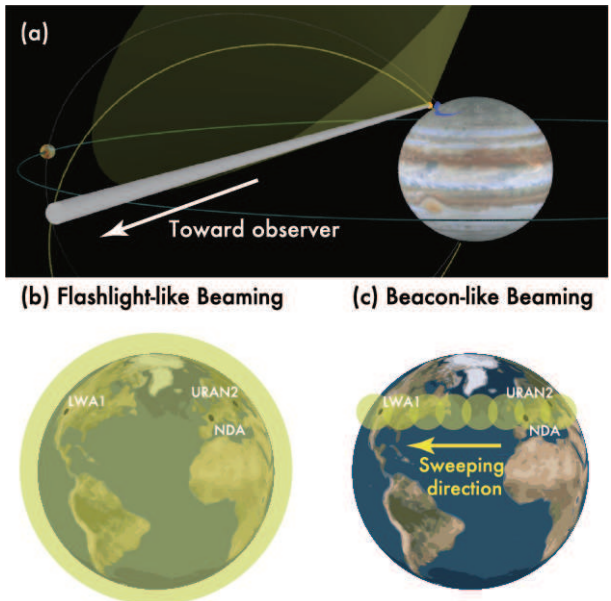}
 \caption{(a) 3D computer graphic (CG) image of Jovian Io-B beaming at Jupiter and schematics of (b) flashlight-like and (c) beacon-like beaming models at Earth. In the case of Io-D, because the generated radio sources are located in southern hemisphere of Jupiter, the beaming geometry is reverse of (a), but (b) and (c) are the exactly same configuration from a view of Earth.}
 \label{fig:beaming}
\end{figure}

In accounting for the observed lag times, we may test the validation of two different beaming models either (1) flashlight-like or (2) beacon-like beaming as drawn in Figure \ref{fig:beaming}. In the case of model 1, the radio sources at Jupiter are located in a wide longitude and isotropically emitted to observers, while model 2 means that the sources are localized in specific longitudes, rotating in longitude about the Jovian magnetic axis. In order to distinguish these two models, it is convenient to compute the propagation differential time vector $\mbox{\boldmath $\tau_\mathrm{prop}$}$ between two stations and the rotation lag time vector $\mbox{\boldmath $\tau_\mathrm{rot}$}$, where each vector composes three independent components from the NDA-URAN2, LWA1-NDA, and LWA1-URAN2 pairs in sequence. Therefore, we may describe the estimated lag time of model 1 ($\mbox{\boldmath $\tau_1$}$) and model 2 ($\mbox{\boldmath $\tau_2$}$) as 
\begin{eqnarray}
\mbox{\boldmath $\tau_1$} &=& \mbox{\boldmath $\tau_\mathrm{prop}$} + \mbox{\boldmath $C$}, \\
\mbox{\boldmath $\tau_2$} &=&  \mbox{\boldmath $\tau_\mathrm{prop}$} +  \mbox{\boldmath $\tau_\mathrm{rot}$} + \mbox{\boldmath $C$}, 
\end{eqnarray}
where $\mbox{\boldmath $C$}$ is the offset lag time vector and, when $\tau_\mathrm{prop}^\mathrm{NU}$ is the propagation differential time of the NDA-URAN2 pair, it is equal to [$-\tau_\mathrm{prop}^\mathrm{NU}$, 15, $15-\tau_\mathrm{prop}^\mathrm{NU}]$ in ms. This is because we approximately corrected the same time at NDA and URAN2 and intentionally inserted positive 15 ms lag time to the LWA1 time stamp. The trend of $\mbox{\boldmath $\tau_1$}$ and $\mbox{\boldmath $\tau_2$}$ are superimposed as the white solid and dotted lines, respectively, in Figure \ref{fig:x-lwa1-nda-uran2}. The white solid lines reasonably fit to the trend of the lag time when the mean cross-correlation functions are relatively high. There is a small scattered area in the initial phase of the LWA1-URAN2 pair in Figure \ref{fig:x-lwa1-nda-uran2}e until 02:04 UT with the angular size of 2.66". This is due to relatively high background noise after the sunset on the LWA1 site. Therefore, it is straightforward to conclude that the beaming thickness has a minimum beaming thickness of 2.66" in the east-west direction projected on Jupiter.

It is important to note that the sharp peak of the mean cross-correlation functions appears at 0 ms lag time because of similar long-scale fluctuation of the background noises (such as artificial interference and Jovian L-bursts) but the occurrence probability of the background noises is less than 4 \% of the total observed time. Nevertheless, these long-scale effects may be easily removed for the S-burst's analysis if we purposely add adequate offset lag time in either pair of telescopes (i.e. our case was positive 15 ms for the LWA1 data).

\begin{figure}
 \centering
  \noindent\includegraphics[width=21pc]{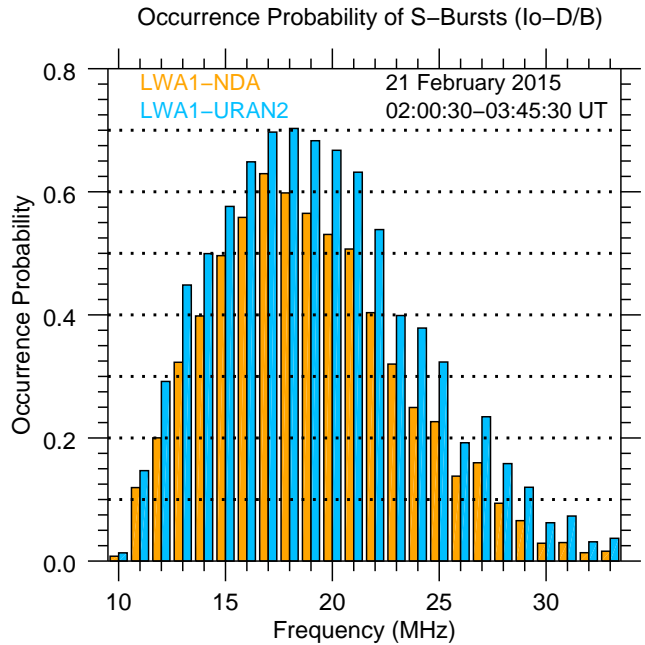}
  \caption{Histogram of occurrence probability of S-bursts within the Io-D/B (i.e. integrated number of occurrences in the surrounded yellow lines of Figure \ref{fig:x-lwa1-nda-uran2}) based on the cross-correlation analysis plotted as a function of frequency.}
 \label{fig:oc_s-bursts}
\end{figure}

In qualitatively estimating the occurrence of S-bursts within the Io-D/B, we integrate a number of occurrences appearing in the surrounded area of yellow lines for $\pm2.5$ ms of the LWA1-NDA and for $\pm5$ ms of the LWA1-URAN2. Figure \ref{fig:oc_s-bursts} shows a plot of the occurrence probability histogram as a function of observed frequency in a pair of LWA1-NDA (orange) and LWA1-URAN2 (blue). The yellow lines in Figure \ref{fig:x-lwa1-nda-uran2} are reflected with the accumulated technical error of timing. The occurrence probability tends to rapidly increase from 10 to 17-18 MHz and then gradually decrease to 33 MHz. This may result from the fact that the low-frequency part corresponds to the Io-D S-bursts and the high-frequency part dominates within the Io-B S-bursts. Conventionally, the discrimination between S- and L-bursts are made by inspecting the observed dynamic spectrum by eye \citep[e.g.][]{Riihimaa:1992ba,Flagg:1991ba,Ryabov:2014aa} or only in the individual S-burst case by computer \citep{Zarka:1996grl}, while our cross-correlation method is simple but efficient to estimate occurrence probability for S-bursts, which includes both simple and complex spectral variation.

\section{Conclusions}
In this paper we report the continuous appearance of S-bursts in a wide frequency band of 10 to 33 MHz over east-west baselines from 0 to 8460 km with the combination of three radio telescopes, namely, the LWA1 in USA, the NDA in France, and the URAN2 in Ukraine. By taking into account the angular size of each pair for the common S-bursts, the beaming tends to come from the same direction of Jupiter with a minimum cone thickness of 2.66". Furthermore, the technique of cross-correlating spectral density obtained from multiple stations used in this study also yields a practical benefit of producing the statistical profiles of S-bursts. We find that the occurrence probability is highest at 17-18 MHz.

Many theoretical studies concerning the generation of Jovian S-bursts consider different wave-particle interactions that occur along the Io flux tube \citep[c.f.][]{Ryabov:2014aa}. In the individual S-burst, the most accepted theory is that a group of upward-moving electrons acts an adiabatic motion along the active magnetic field lines near Io, thereby converting released energy into S-bursts via the electron cyclotron maser instability \citep{Zarka:1996grl}. Our result from one observing direction implies two kinds of possible scenarios of S-burst mechanism. First case is the short duration of radiation having isotropic wide beaming. Second case is the long duration of radio emission with a confined narrow beam, thereby suggesting that the beam size depends upon an observed event's duration and the rotation period of the bodies of interest such as Io and Jupiter. By comparing the measurements of the wavefront arrival times from several observing directions, the ambiguity of two cases can be solved. Thus, our proposed cross-correlation method applied to more combined and extended observations with distant telescopes, may provide an insight into a better understanding of the S-burst mechanism. 

\acknowledgments

The authors are pleased to acknowledge the engineers and staff who operate LWA1, NDA, and URAN2 radio telescopes. Construction of the LWA1 was supported by the Office of Naval Research under Contract N00014-07-C-0147. Support for operations and continuing development of the LWA1 is provided by the National Science Foundation under grants AST-1139963 and AST-1139974 of the University Radio Observatory program. Basic research in radio astronomy at the Naval Research Laboratory is supported by 6.1 Base funding. Part of this research was performed while one of the authors (M.~I.) was on leave at LESIA of the Observatoire de Paris-Meudon. M.~I.~acknowledges support from a Grant-in-Aid (13J01567) for Research Fellows of the Japan Society for the Promotion of Science (JSPS), and also thanks LESIA of the Observatoire de Paris-Meudon for the generous hospitality during his one-year visit. The work of M.~P.~was supported by the Austrian Science Fund (project P23762-N16). This research has been supported in part by JSPS KAKENHI grant 25400480.

\end{document}